\documentclass[a4paper]{jpconf}
\usepackage{cite}
\usepackage{graphicx}
\usepackage{epsfig}
\newcommand{\e}[1]{\epsilon_{#1}}
\newcommand{\newc}{\newcommand}
\newc{\ra}{\rightarrow}
\newc{\lra}{\leftrightarrow}
\def\tila{\tilde{\alpha}}

\newcommand{\ba}{\begin{eqnarray}}
\newcommand{\ea}{\end{eqnarray}}
\newcommand{\ppr}{{\it Preprint} }
\newcommand{\JHEP}{{\it JHEP} }
\begin{document}
\title{Unification and fermion mass relations in low string scale D-brane models}
\author{D V Gioutsos\footnote[1]{Talk presented at the ``Corfu Summer Institute'',
Corfu-Greece, September 4-14, 2005. Work done in colla\-boration with
G~K~Leontaris and J~Rizos.}}
\address{Theoretical Physics Division, Ioannina University, GR-45110 Ioannina, Greece}
\begin{abstract}
In this talk, gauge coupling evolution is analyzed in D-brane inspired models
with two Higgs doublets and a $U(3) \times U(2) \times U(1)^N$ gauge symmetry. In
particular, we focus on D-brane configurations with two or three abelian factors.
We  find that the correct hypercharge assignment of the Standard Model particles
is reproduced for six viable models distinguished by different brane
configurations. We also investigate the bottom tau quark mass relation and find
that the correct low energy $m_b / m_\tau$ ratio is obtained for equal $b-\tau$
Yukawa couplings at a string scale around $10^3$ TeV.
\end{abstract}

\section{Introduction}
Low scale unification of gauge and gravitational interactions
~\cite{Antoniadis:1990ew,Arkani-Hamed:1998rs,Antoniadis:1998ig}, appears to be a
promising framework for solving the hierarchy problem. In this context, the
weakness of the gravitational force in long distances is attributed to the
existence of extra dimensions at the Fermi scale. A realization of this scenario
can occur in type I string theory~\cite{Lykken:1996fj} where gauge interactions
are mediated by open strings with their ends attached on some D-brane stack,
while gravity is mediated by closed strings that propagate in the whole 10
dimensional space.

In the context of Type I string theory using appropriate collections of parallel
\cite{Polchinski:1996na,Angelantonj:2002ct} or intersecting
\cite{Berkooz:1996km,Balasubramanian:1996uc} D-branes, there has been
considerable work in trying to derive the Standard Model theory or its Grand
Unified extensions \cite{Antoniadis:2002en, Antoniadis:2002qm, Antoniadis:2002cs,
Gioutsos:2005uw, Aldazabal:2000cn, Cvetic:2001tj, Leontaris:2001hh, Kokorelis,
Blumenhagen:2003jy, Blumenhagen:2005mu, Antoniadis:2004dt}. Some of these low
energy models revealed rather interesting features: (i) The correct value of the
weak mixing angle is obtained for a string scale of the order of a few TeV (ii)
baryon and lepton numbers are conserved due to the existence of  exact global
symmetries which are remnants of additional anomalous $U(1)$ factors broken by
the Green-Schwarz mechanism (iii) supersymmetry is not necessary  to solve the
hierarchy problem.

However, its rivals,  supersymmetric Grand Unified theories (where the
unification of gauge couplings occurs at the order of $10^{16}$ GeV), and their
heterotic string realizations (with even higher unification scale), exhibit also
a number of additional interesting features.  Apart from the  natural gauge
coupling unification these features include fermion mass~\cite{Chanowitz:1977ye,
Greene:1986jb} relations and in particular the bottom tau-unification, i.e. the
equality of the corresponding Yukawa couplings at the unification scale, which
reproduces the correct mass relation at low energies.

Full gauge coupling unification does not occur in low string scale models,
however, this should not be considered as a drawback since the various gauge
group factors are associated with different stacks of branes and therefore gauge
couplings may  differ at the string scale. In standard-like models in particular,
there should be at least three different stacks of branes accommodating the
$SU(3)$, $SU(2)$ and $U(1)$ gauge groups respectively.

Following a bottom-up approach~\cite{Gioutsos:2005uw}, in this talk  we examine
the possible brane configurations that can accommodate the Standard Model and the
associated hypercharge embeddings and we analyze the consequences of (partial)
gauge coupling unification in conjunction with bottom-tau Yukawa coupling
equality. We shall restrict to non-super\-symmetric configurations, (for some
recent results on supersymmetric and split supersymmetric models see
\cite{Blumenhagen:2003jy, Antoniadis:2004dt} and references therein), however, we
will consider models with two Higgs doublets so that the bottom and top quark
masses will be related to different vacuum expectation values while the tau
lepton and the bottom quark will  receive masses from the same Higgs doublet. We
find that in a class of models that can be realized in the context of type I
string theory with large extra dimensions, the experimentally low energy masses
can be reproduced assuming equality of bottom-tau Yukawa couplings and a string
scale as low as $\sim 10^3$ TeV.

In the next section we briefly describe the general set up of brane models and
derive the hypercharge formulae for an arbitrary number of $U(1)$ factors. In
section 3 we identify two brane configurations that admit only one Higgs doublet
coupled to the down quarks and leptons: the first is a four brane-stack
configuration with two $U(1)$ branes, while the second is a five brane-stack
system with three $U(1)$ branes. Section 4 deals with the calculational details
and renormalization analysis of gauge couplings, while in section 5 the results
for $b-\tau$ Yukawa couplings are presented. Our conclusions are drawn in section
6.

\section{Hypercharge embedding in generic Standard model like brane configurations}
We consider models which arise in the context of various D-brane configurations
\cite{Antoniadis:2002en, Antoniadis:2002qm}. A single D-brane carries a $U(1)$
gauge symmetry which is the result of the reduction of the ten-dimensional
Yang-Mills theory. Therefore, a stack of $n$ parallel D-branes gives rise to a
$U(n)$ gauge theory where the gauge bosons correspond to open strings having both
their ends attached to some of the branes of the various stacks.

The minimal number of brane sets  required to provide the Standard Model
structure is three: a 3-brane ``color" stack with gauge symmetry
${U(3)}_C\sim{SU(3)}_C\times{U(1)}$, a 2-brane ``weak" stack which gives rise to
${U(2)}_L\sim{SU(2)}_L\times{U(1)}$ gauge symmetry and an abelian $U(1)$ brane
for hypercharge. However, accommodation of  all SM particles as open strings
between different brane sets requires at least one $U(1)$ brane to be added to
the above configuration~\cite{Antoniadis:2002en,Antoniadis:2002cs}. Additional
abelian branes may be present too. In more complicated scenarios the weak or
color stacks can be repeated leading to an effective ``higher level embedding" of
the Standard Model. The full gauge group will be of the form%
\ba
G={U(m)}_C^p\times{U(n)}_L^q\times{U(1)}^N \label{ggg}
\ea%
with $m\ge 3$ and $n\ge 2$ and $p,q\ge1$. Since $U(n)\sim {SU(n)}\times{U(1)}$
and so on, we infer that brane constructions automatically give rise to models
with $SU(n)$ gauge group structure and several $U(1)$ factors.

A generic feature of this type of string vacua is that several abelian gauge
factors are anomalous. However, at least one $U(1)$ combination  remains anomaly
free. This is the hypercharge that can be in general written as%
\ba
Y=\sum_{i=1}^p k_3^{(i)} Q_3^i + \sum_{j=1}^q k_2^{(j)} Q_2^j+
\sum_{\ell=1}^N k'_\ell \, Q'_\ell, \label{ydef}
\ea%
where $Q_3^i$ are the $U(1)$ generators of the color factor $i$, $Q_2^j$ are the
${U(1)}$ generators of the weak factor $j$ and $Q'_\ell\,(\ell=1,\dots,N)$, are
the generators of the remaining Abelian factors.

The  simplest case which leads directly to the SM theory is the choice $p=q=1$.
Constructions  of this type  have already been proposed in
reference~\cite{Antoniadis:2002en}. An immediate consequence of (\ref{ggg}) and
(\ref{ydef})  is that the hypercharge coupling ($g_Y$) at the string/brane scale
$(M_S)$ is related to the brane couplings ($g_m, g_n, g_i'$) as%
\ba
\frac{1}{g_Y^2}=\frac{2 m k_3^2}{g_m^2}+\frac{2 n k_2^2}{g_n^2}+2\sum_{i=1}^N
\frac{{k'_i}^2}{ {g'_i}^2}\label{gydef}
\ea%
where we have used the traditional normalization ${\rm Tr}\, T^a\, T^b=
\delta^{ab}/2 , a,b=1, \dots, n^2$ for the $U(n)$ generators and assumed that the
vector representation (${\bf n}$) has abelian charge $+1$ and thus the $U(1)$
coupling becomes ${g_n}/{\sqrt{2 n}}$ where $g_n$ the $SU(n)$ coupling.

Choosing further $m=3, n=2$ in (\ref{gydef}) we obtain directly the non-abelian
structure of the SM with several $U(1)$ factors, therefore the hypercharge gauge
coupling condition reads%
\ba
 k_Y&\equiv&\frac{\alpha_2}{\alpha_Y}
   \;=\;{6 k_3^2}\,\frac{\alpha_2}{\alpha_3}+4 k_2^2+2\sum_{i=1}^N
{k_i'}^2\,\frac{\alpha_2}{\alpha_i'}\label{kY}
\ea%
where $\alpha_i \equiv g_i^2/(4\pi)$. Given a relation between the $\alpha_i'$
and $\alpha_2$ (or $\alpha_3$) and a hypercharge embedding ($k_i'$ known)
equation (\ref{kY}) in conjunction with the $\alpha_3$ evolution equation,
determine the string scale $M_S$. In the remaining of this section,  we will
derive  all possible sets of $k_i$'s compatible with brane configurations which
embed the SM particles and imply an economical Higgs spectrum.

\begin{figure}[!b]
\centering ($N=2$) \includegraphics[width=0.28\textwidth]{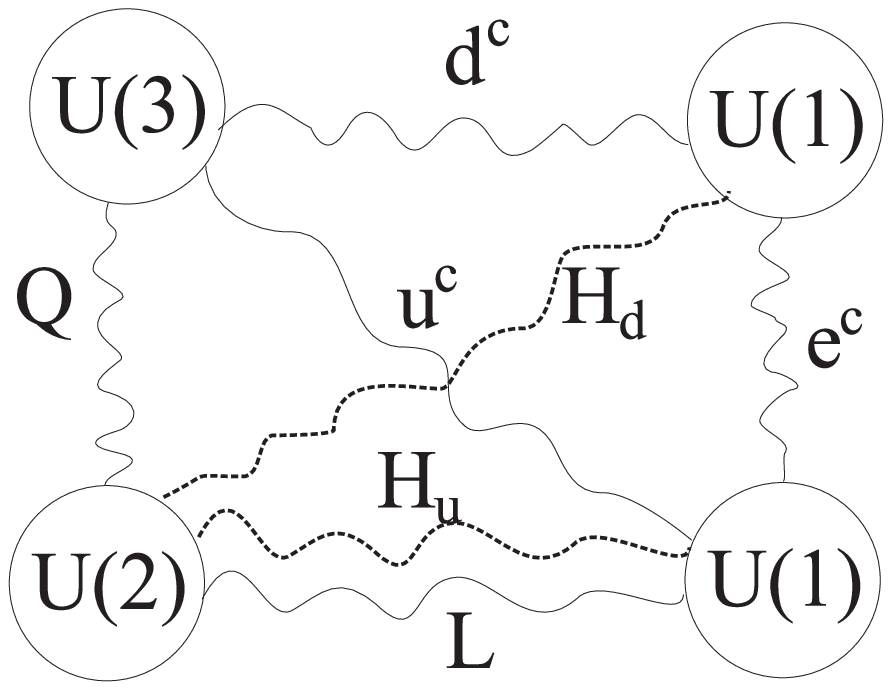}\ \ \ \ \ \
\ \ \ \ \ \ \ \ \ ($N=3$)
\includegraphics[width=0.25\textwidth]{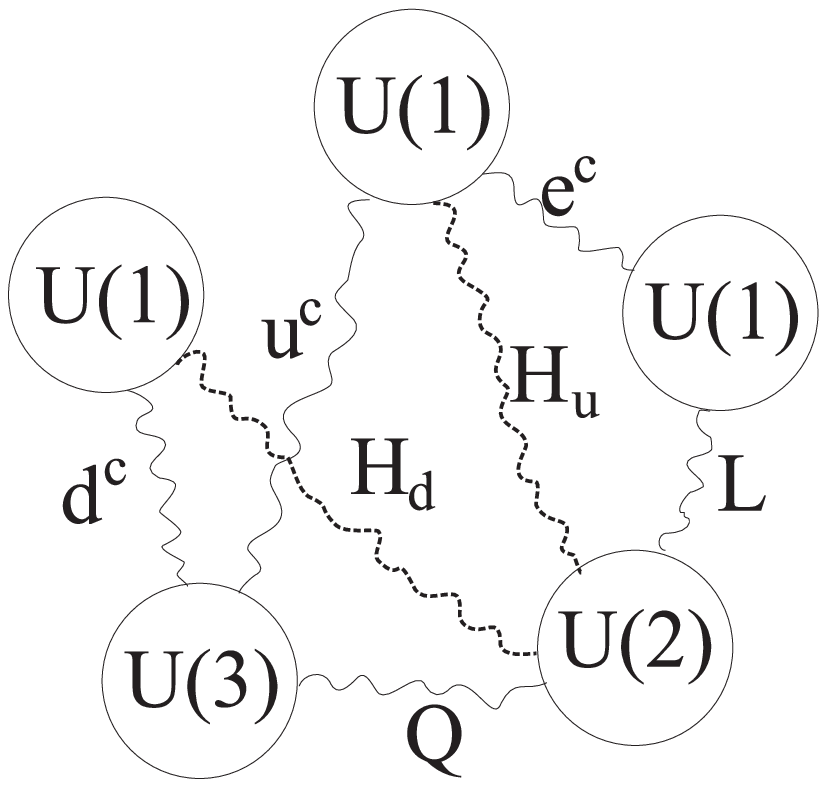}
\caption{\label{fconf} Possible $N=2,3$ brane configurations ($N$ is the number
of the $U(1)$-branes) that can accommodate the SM spectrum with down quarks and
leptons acquiring masses from the same Higgs doublet.}
\end{figure}
\section{Concrete brane configurations\label{cbc}}
We consider here  brane configurations that can lead to concrete realizations of
those proposed previously. As already mentioned, a specific realization must
include two Higgs doublets in order  to ensure the bottom-top mass difference. In
brane models each SM particle  corresponds to an open string stretched between
two branes. In our charge conventions, the possible quantum numbers of such a
string ending to the $U(m)$ and $U(n)$ brane sets are $\left({\bf m};+1,{\bf
n};+1\right),\left({\bf \bar{m}};-1,{\bf n};+1\right)$, $\left({\bf m};+1,{\bf
\bar{n}};-1\right)$, $\left({\bf \bar{m}};-1,{\bf \bar{n}}; -1 \right)$, that is,
bifundamentals of the associated unitary groups. Higher representations could be
obtained by considering strings with both ends on the  same brane set $U(m)$,
$\left({\bf m(m-1)/2}, 2 \right)$, $\left({\bf m(m+1)/2}, 2 \right)$, $\dots$,
however,  we will restrict here to the bi-fundamental case. By analyzing possible
brane configurations that can accommodate  a gauge group of the form (\ref{ggg})
we find that only the four and five brane-stack scenarios ($N=2,3$) in
(\ref{ggg}) can lead to natural $b$-$\tau$ unification. It is possible to
introduce additional brane sets, however, in such a case down quarks and leptons
get  their masses from different Higgs doublets and any Yukawa coupling
unification condition would require the equality of the associated doublet vevs.
The $N=2,3$ two Higgs doublet candidate configurations are presented pictorially
in figure \ref{fconf}.

The associated hypercharge embeddings  can be obtained by solving the hypercharge
assignment conditions for SM particles for $k_i$. SM particle abelian charges
under $U_3(1) \times U_2(1) \times {U(1)}_1 \times {U(1)}_2$ are the general form
$Q(+1,\epsilon_1,0,0)$, $d^c(-1,0,\epsilon_2,0)$, $u^c(-1,0,0,\epsilon_3)$,
$L(0,\epsilon_4,0,\epsilon_5)$, $e^c(0,0,\epsilon_6,\epsilon_7)$ and thus%
\ba
k_3+k_2\, \e1 &=& \hphantom{+}\frac{1}{6}\nonumber\\
-k_3+k_1'\, \e2 &=& \hphantom{+}\frac{1}{3}\nonumber\\
-k_3+k\, \e3 &=& -\frac{2}{3}\label{6e}\\
k_2\, \e4+k_2'\, \e5 &=& -\frac{1}{2}\nonumber\\
k_1'\,\e6 +k_2'\, \e7 &=& \hphantom{+}1\nonumber
\ea%
where $k=k_2'$ in the first configuration and $k=k_3'$ for the second one, while
$\e{i}^2=1,\,i=1,\dots,7$. As seen by  (\ref{ydef}) and (\ref{gydef}), only the
absolute values of the hypercharge embedding coefficients $k_i,\,k_i'$ enter the
coupling relation at $M_S$. Solving (\ref{6e}), for the SM particle charges in
configuration ($N=2$) we obtain three possible solutions. These correspond to the
(absolute) values for the coefficients presented in cases (a), (b) and (c) of
table \ref{ytab1}. Configuration $N=3$ leads to four additional cases, namely
(d), (e), (f) and (g) of the same table. If in a particular solution a
coefficient $k_i$ (or $k_i'$) turns out to be zero, the associated abelian factor
does not participate to the hypercharge.
\begin{table}[!t]
\caption{\label{ytab1}Absolute values of the possible hypercharge embedding
coefficient sets ($k_3, k_2$ and $k_i'$)  for the brane configurations with $N=2$
and $N=3$ of figure \ref{fconf}.}
\renewcommand{\arraystretch}{1.2}
\centering
\begin{tabular}{clccccc}
\br
$N$&&$|k_3|$&$|k_2|$&$|k_1'|$&$|k_2'|$&$|k_3'|$\\
\mr
&(a)&$\frac{1}{6}$&$0$&$\frac{1}{2}$&$\frac{1}{2}$&-\\
$2$&(b)&$\frac{2}{3}$&$\frac{1}{2}$&$1$&$0$&-\\
&(c)&$\frac{1}{3}$&$\frac{1}{2}$&$0$&$1$&-\\
\mr
&(d)&$\frac{1}{6}$&$0$&$\frac{1}{2}$&$\frac{1}{2}$&$\frac{1}{2}$\\
&(e)&$\frac{1}{3}$&$\frac{1}{2}$&$0$&$1$&$1$\\
$3$&(f)&$\frac{5}{6}$&$1$&$\frac{1}{2}$&$\frac{1}{2}$&$\frac{3}{2}$\\
&(g)&$\frac{2}{3}$&$\frac{1}{2}$&$1$&$0$&0\\
\br
\end{tabular}

\end{table}

\section{Gauge coupling running and the String scale}
Following a bottom-up approach, in this section we determine the range of the
string scale for all the above models by taking into account the experimental
values of $\alpha_3, \alpha_e$ and $\sin^2\theta_W$ at $M_Z$ \cite{Eidelman}
 \ba
\alpha_3 = 0.118\pm 0.003,~~~\alpha^{-1}_{e}=127.906,~~~
\sin^2\theta_W=0.23120\nonumber \ea For the scales above $M_Z$ we consider the
standard model spectrum with two Higgs doublets. The one loop RGEs  for the gauge
couplings ($\tilde{\alpha}\equiv \alpha/(4\pi)$) take the form \ba \frac{d
\tilde{\alpha}_i}{dt} = b_i \tilde{\alpha}_i^2\,,~~~~ i=Y,2,3\label{rge} \ea
where $(b_Y, b_2, b_3)= (7, -3, -7)$ and $t=2\ln\mu$ ($\mu$ is the
renormalization point).
\begin{table}[!t]
\caption{\label{ytab2} Possible values of $k_Y$ as a function of $\xi=\alpha_2 /
\alpha_3$ for various orientations of $U(1)$'s for the models of table
\ref{ytab1}. The rows show the $k_Y$ values for various orientations (see text
for details). Last row shows the minimum value of the string scale $M_S$ obtained
for the models (a)-(g).}%
\renewcommand{\arraystretch}{1.2}
\centering
\begin{tabular}{ccccccc}
\br
&\multicolumn{6}{c}{Model}\\
\mr%
\parbox{1.5cm}{\small coupling\\[-4pt]relation}&(a)&(b)&(c)&(d)&(e)&(g)\\
\mr%
 $\alpha_i'=\alpha_2$&$\frac{\xi}{6}+1$&$\frac{8\xi}{3}+3$ &
$\frac{2\xi}{3}+3$ & $\frac{\xi}{6}+\frac{3}{2}$ & $\frac{2\xi}{3}+5$ & $\frac{8\xi}{3}+3$ \\
$\alpha_i'=\alpha_3$&$\frac{7\xi}{6}$ & $\frac{1\bf{4\xi}}{3}+1$ &
$\frac{8\xi}{3}+1$ &
$\frac{2\xi}{3}+1$ & $\frac{14\xi}{3}+1$ & $\frac{14\xi}{3}+1$ \\
(see text)&$\frac{2\xi}{3} +\frac{1}{2}$ & - & - & $\frac{5\xi}{3}$ & $\frac{8\xi}3+3$ & - \\
(see text)&- & - & - & $\frac{7\xi}{6}+\frac 12$ & - & - \\
\mr
$M_S $(GeV)&$6.71\times 10^{17}$&$5.78\times 10^3$&$1.99\times
10^6$&$1.65\times 10^{14}$
&$5.78\times 10^3$&$5.78\times 10^3$\\
\br
\end{tabular}
\end{table}

First, we concentrate on simple relations of the gauge couplings, i.e., those
relations implied from  models arising only in the context on non-intersecting
branes.  In these cases, certain constraints on the initial values of the gauge
couplings have to be taken into account, leading to a discrete number of
admissible cases which we are going to discuss. Thus,  in the case of two $D5$
branes, $U(3)$ and $U(2)$ are confined in different bulk directions. In the
parallel brane scenario the orientation of a number of the extra $U(1)$'s may
coincide with the $U(3)$-stack direction while the remaining abelian branes are
parallel to the $U(2)$ stack. This implies that the corresponding $U(1)$ gauge
couplings have the same initial values either with the $\alpha_3$ or with the
$\alpha_2$ gauge couplings. If we define $\xi=\frac{\alpha_2}{\alpha_3}$ the
ratio of the two non-abelian gauge couplings at the string scale, for any
distinct case,  $k_Y$ takes the form $k_Y= \lambda \,\xi +\nu$, where
$\lambda,\nu$ are calculable coefficients which depend on the specific
orientation of the $U(1)$ branes. For example, in model (a) we can have the
following possibilities: $\alpha_1'=\alpha_2'=\alpha_2$,
$\alpha_1'=\alpha_2'=\alpha_3$ and $\alpha_1'=\alpha_2, \alpha_2'=\alpha_3$
leading to $k_Y=\frac{\xi}6+1$, $\frac{7\xi}6$ and $\frac{2\xi}3+\frac 12$
correspondingly. All cases for the models (a)-(g)  are presented in
table~\ref{ytab2} and are classified with regard to the hypercharge coefficient
$k_Y$. (All cases of Model (f) lead to unacceptably small string scales, so these
are not presented).

Allowing ${\alpha_3}$ to take values different from ${\alpha_2}$, we find that
models (a,b,c,d,e,g) of table \ref{ytab1} predict a string scale in a wide range,
from a few TeV up to the Planck mass. The highest value  is of the order $M_S\sim
7\times 10^{17}$ GeV and corresponds to equal couplings
$\frac{\alpha_2}{\alpha_3}\equiv \xi =1$ at $M_S$. On the other hand, lower
unification values of the order of a few  TeV assume a gauge coupling ratio
$\frac{\alpha_3}{\alpha_2}\approx 2$. In this case the idea of complete gauge
coupling unification could be still valid,  considering that the SM gauge group
arises from the breaking of a gauge symmetry whose non-abelian part is
$U(3)\times U(2)^2$, i.e., for the case $p=1,\,q=2$ of (\ref{ggg}) where the
factor of 2 in the gauge coupling ratio is related to the diagonal breaking
$U(2)\times U(2)\ra U(2)$. The lowest possible unification for the three models
(b),(e),(g) corresponds to  $k_Y=\frac{14\,\xi}3+1$, and is $M_S\sim 5.81\times
10^{3}$ GeV, for a weak to strong gauge coupling ratio $\xi\sim 0.42$ at $M_S$.
Case (c) predicts an intermediate value $M_S =2 \times 10^6$ GeV while model (d)
gives $M_S\sim 10^{14}$ GeV. Finally, model (a) for $\xi\sim 1$ predicts a
unification scale as high as $M_S\sim 6.7\times 10^{17}$ GeV which is of the
order of the heterotic string scale. Interestingly, in this latter case, all
gauge couplings are equal at $M_S$, $\alpha_3=\alpha_2=\alpha_i'$, while, as can
be seen from table 2, $k_Y$ takes a common value for all three cases, $k_Y=7/6$.
\begin{figure}[!t]
\hspace*{-.5cm}
\includegraphics[width=0.5\textwidth]{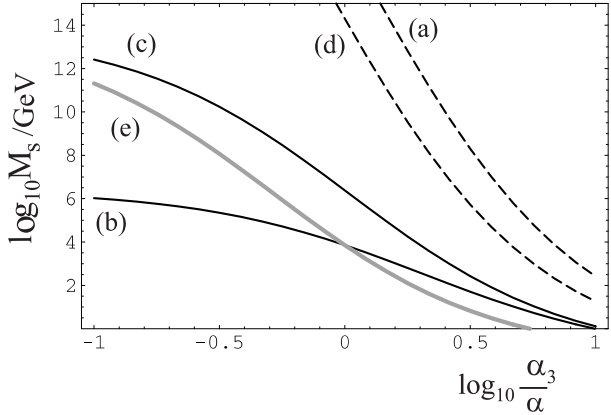}\hspace{1.0pc}%
\begin{minipage}[b]{18pc}
\caption{\label{cmodels}The string scale as a function of the coupling ratio
$\frac{\alpha_3}{\alpha}$, ($\alpha$ is a common value for the $U(1)$ couplings
$\alpha_i'$) for the hypercharge embeddings of table 1, in the general case of
intersecting branes. Re\-sults for model (g) coincide with those of model
(b).}\vspace*{0.2cm}
\end{minipage}
\end{figure}

In the general intersecting case, the $U(1)$ branes are neither aligned to the
$SU(3)$, nor to the $SU(2)$ stacks, thus the corresponding gauge couplings can
take arbitrary values.  Without loss of generality, we will assume here for
simplicity that all these couplings are equal
$\alpha_1'=\alpha_2'=\dots=\alpha_N'=\alpha$. In figure \ref{cmodels} we plot the
string scale ($M_S$) as a function of the logarithm of the ratio
$\alpha_3/\alpha$ for the candidate models (a), (b), (c), (d), (e) and (g).  The
results for models (b), (c), (e) and (g) which is identical with model (b), are
represented in the figure with continuous lines. These are compatible with low
scale unification particularly when $\alpha_3\ge \alpha$. For
$\alpha_i'=\alpha_3$, (which corresponds to the zero of the logarithm at the
$x$-axis), we obtain again the results of the parallel brane scenario, shown in
table 2. At this point, we further observe a crossing of the (e)-curve with the
curve for models (b),(g). It is exactly this point ($\alpha_i'=\alpha_3$) that
these three models predict the same value for the lowest string scale. When
$\alpha_3\ge \alpha_i'$, model (e) predicts the lowest $M_S$, whilst, if
$\alpha_i'>\alpha_3$, models (b), (g) imply lower string scales than model (e).

The values of the string scale for models (a), (d) (represented in the figure
with dashed curves) are substantially higher; for these latter cases in
particular, assuming reasonable gauge coupling relations $\alpha_i'\approx {\cal
O}( \alpha_{2,3})$  we find that $M_S\ge 10^{12}$ GeV. Again, for
$\alpha_3=\alpha_i'$, (the zero value of the $x$-axis) we rederive the values of
$M_S$ presented in table 2.

\section{Yukawa coupling evolution and mass relations}
In this section, we will examine whether a unifiaction  of the $b - \tau$ Yukawa
couplings\footnote{%
For $b-\tau$ unification in a different context see also~\cite{Parida:1996mz}.} %
is possible in the above described low string scale models. Our procedure is the
following: Using the experimentally determined values for the third generation
fermion masses $m_b, m_{\tau}$ we run the 2-loop system of the $SU(3)_C\times
U(1)_Y$ renormalization group equations up to the weak scale ($M_Z$) and
reconcile there the results with the experimentally known values for the weak
mixing angle and the gauge couplings. For the renormalization group running below
$M_Z$ we define the parameters
\begin{eqnarray}
\tila_e \;=\; \left( \frac{e}{4 \pi} \right)^2 ,\, \tila_3 \;=\; \left(
\frac{g_3}{4 \pi} \right)^2 ,\, t \;=\; 2 \ln\mu
\end{eqnarray}
where $e, g_3$ are the electromagnetic and strong couplings respectively and
$\mu$ is the renormalization scale. The relevant
RGEs are \cite{Arason:1991ic}%
\begin{eqnarray}
\frac{d\tila_e}{dt} &=& \frac{80}{9}\tila_e^2+
\frac{464}{27}\tila_e^3+\frac{176}{9}\tila_e^2\tila_3\nonumber
\\[2mm]
\frac{d\tila_3}{dt} &=& -\frac{23}{3}\tila_3^2 - \frac{116}{3}\tila_3^3 +
\frac{22}{9}\tila_3^2\tila_e - \frac{9769}{54}\tila_3^4 \nonumber
\\[2mm]
\frac{dm_b}{dt} &=& m_b \left\{ -\frac{1}{3} \tila_e - 4\tila_3 +
\frac{397}{162}\tila_e^2 - \frac{1012}{18} \tila_3^2 - \frac{4}{9} \tila_3
\tila_e - 474.8712 \tila_3^3 \right\}\nonumber
\\[2mm]
\frac{dm_\tau}{dt} &=& m_\tau \left\{ -3\tila_e + \frac{373}{18}\tila_e^2
\right\}\nonumber
\end{eqnarray}
where $m_b, m_\tau$ are the running masses of the bottom quark and the tau lepton
respectively, while we use the notation $\tilde{a}_i \equiv g^2_i/16\pi^2$ and
$\tilde{a}_{t,b,\tau} \equiv \lambda^2_{t,b,\tau}/16\pi^2$.

The required value for the running mass of $m_t$ at $M_Z$ is computed as follows:
we formally solve the 1-loop RGE system for ($\tilde{a}_3$, $\tilde{a}_2$,
$\tilde{a}_Y$, $\tilde{a}_t$, $\tilde{a}_b$, $\tilde{a}_\tau$) and afterwards we
determine the interpolating function for $\tilde{a}_3(\mu)$ and $m_t(\mu; m_t^z)$
at any scale $\mu$ above $M_Z$, where $m_t^z \equiv m_t(M_Z)$ indicates the
dependence on an arbitrary initial condition. The unknown value for $m_t^z$ is
determined by solving numerically the algebraic equation%
\ba%
\left[ m_t(\mu; m_t^z) - \frac{M_t}{1+ \frac{16}{3} \tilde{a}_3(\mu)-2
\tilde{a}_t(\mu)} \right]_{\mu=M_t}=0
\ea%
We use these results as inputs for the relevant parameters and we run the RGE
system to higher scales until the $\tilde{a}_b$ and $\tilde{a}_{\tau}$ Yukawa
couplings coincide. The scale that this happens is considered as the string
scale. There, the values of $\tilde{a}_3, \tilde{a}_2, \tilde{a}_Y$ are checked
and the ratio $\tilde{a}_2/\tilde{a}_Y$ is calculated in order to obtain the
normalization constant $k_Y$. In our numerical analysis we use for the gauge
couplings the values presented in the previous section, for the bottom quark mass
$m_b$ the experimentally determined range at the scale  $\mu = m_b$ ie.~$m_b(m_b)
= 4.25 \pm 0.15$ GeV and finally the top pole mass is taken to be $M_t=178.0 \pm
4.3$ GeV \cite{Eidelman}.

For the scales above $M_Z$ we consider the standard model spectrum augmented by
one more Higgs. The Higgs doubling is in accordance with the situation that
usually arises in the SM variants with brane origin. Moreover, we assume that one
Higgs $H_u$ only couples to the top quark while the second Higgs $H_d$ couples
only to the bottom. Then, in analogy with supersymmetry we define the angle
$\beta$ related to their vevs where $\tan\beta=\frac{v_u}{v_d}$. Thus, we have
the equations for the gauge couplings
\begin{eqnarray}
  \frac{d\tila_Y}{dt}\;=\;  7 \tila_Y^2 ,\;\;
  \frac{d\tila_2}{dt}\;=\; -3 \tila_2^2 ,\;\;
  \frac{d\tila_3}{dt}\;=\; -7 \tila_3^2 \nonumber
  \end{eqnarray}
  and for the Yukawas
  \begin{eqnarray}
  \frac{d\tila_t}{dt}&=&  \tila_t (-\frac{17}{12}\tila_Y
  - \frac{9}{4} \tila_2 - 8 \tila_3 + \frac{9}{2} \tila_t + \frac{1}{2} \tila_b)  \nonumber\\
  \frac{d\tila_b}{dt} &=& \tila_b (-\frac{5}{12} \tila_Y
  - \frac{9}{4} \tila_2 -8 \tila_3 + \frac{1}{2} \tila_t + \frac{9}{2} \tila_b + \tila_{\tau}) \nonumber\\
  \frac{d\tila_{\tau}}{dt} &=&  \tila_{\tau} (-\frac{15}{4} \tila_Y
  - \frac{9}{4} \tila_2 + 3 \tila_b + \frac{5}{2} \tila_{\tau} )   \nonumber\\
  \frac{dv_u}{dt} &=& \frac{v_u}{2} (\frac{3}{4} \tila_Y + \frac{9}{4} \tila_2
  - 3 \tila_t)           \nonumber   \\
  \frac{dv_d}{dt} &=& \frac{v_d}{2} (\frac{3}{4} \tila_Y + \frac{9}{4} \tila_2
  - 3 \tila_b - \tila_{\tau} )\nonumber
\end{eqnarray}
where  $ t = 2\ln\mu$.

Further, if we define $ v^2 = v_u^2 +v_d^2 $, with $v_u = v \sin\beta $, $v_d = v
\cos\beta$ and $v\sim 174~ {\rm GeV}$, the $Z$-boson mass is given by $M_Z^2 =
\frac{1}{2}(g_Y^2 + g_2^2)v^2$. The elecromagnetic and the strong couplings are
defined in the usual way
\[ \tilde{\alpha}_{e} =  \tilde{\alpha}_Y \cos^2\theta_W =
\tilde{\alpha}_2 \sin^2\theta_W \] while the top and bottom quark masses are
related to the Higgs vevs by
\[ m_t = 4 \pi v_u \sqrt{\tilde{\alpha}_t} ~~~~~~~~ m_b = 4 \pi v_d \sqrt{\tilde{\alpha}_b} \]
\begin{table}[!t]
\caption{\label{ytab3a} The  String scale and the $b-\tau$ ratio at $M_S$ for
various orientations of $U(1)$ branes presented in table \ref{ytab2}.}%
\centering
\begin{tabular}{ccccc}
\br
model & $k_Y$ & $\xi=\frac{\alpha_2}{\alpha_3}$ & ${M_S}/{GeV}$&$\frac{m_b}{m_{\tau}}(M_S)$\\
\mr
b,e,g & $2.969$ & 0.42 & $5.786 \times 10^3$ &1.25     \\
c     & $2.539$ & 0.58 & $1.986 \times 10^6$  &1.01    \\
d     & $1.554$ & 0.93 & $1.645 \times 10^{14}$  &0.73  \\
a     & $1.226$ & 1.01 & $6.710 \times 10^{17}$ &0.68   \\
\br
\end{tabular}
\end{table}

We will examine the possibility of obtaining $b-\tau$ unification at a low string
scale $M_S$. We first concentrate in the models (a)-(g) discussed in the previews
section. We present our results in the last column of table \ref{ytab3a}. We
notice that $b-\tau$ unification is obtained in model $c$, for $M_S\approx 10^6$
GeV. Models (b), (e), (g) with unification scale $M_S\approx 5.8 \times 10^3$ GeV
predict a small deviation from exact $b-\tau$ unification. We observe that in
these cases the strong-weak gauge coupling ratio\footnote{This relation holds
naturally if we embed the model in a $U(3)\times U(2)^2\times U(1)^2$ symmetry.}
is ${a_3}\approx 2\,{a_2}$.
\begin{figure}[!h]
\hspace*{-0.9cm}
\includegraphics[width=0.6\textwidth]{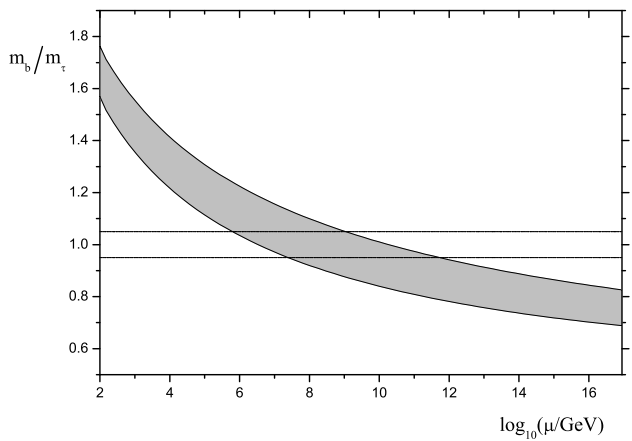}\hspace{-1pc}%
\begin{minipage}[b]{18pc}
\caption{\label{mbmtau}The ratio $\frac{m_b}{m_{\tau}}$ as a function of the
energy $\mu$ in the 2-Higgs Standard Model. The shaded region corresponds to
$a_3$ and threshold uncertainties.}\vspace*{0.7cm}
\end{minipage}
\end{figure}

In figure \ref{mbmtau} the ratio $m_b/m_\tau$ is plotted as a function of the
energy scale for the case of the two-Higgs Standard Model (see~\cite{Kane:2005va}
and references therein). All previous uncertainties are incorporated and the
result is the shaded region shown in the figure. The horizontal shaded band is
defined between the values $\frac{m_b}{m_{\tau}}=[0.95-1.05]$ and take into
account deviations of the ratio $\frac{m_b}{m_{\tau}}$ from unity due to possible
threshold as well as mixing effects in the full $3\times 3$ quark and lepton
flavor mass matrices. As can be seen, exact $m_b=m_{\tau}$ equality is found
around the scale $M_S\approx 10^6$ GeV. Taking into consideration
$m_b/m_{\tau}$-uncertainties expressed through the shaded band, the $M_S$ energy
range is extended up to $\sim 10^{12}$ GeV.

\section{Conclusions}

We performed a systematic study of the Standard Model embedding in brane
configurations with $U(3)\times U(2)\times U(1)^N$ gauge symmetry and we examined
a number of interesting phenomenological issues. Seeking for models with
economical Higgs sector, we identified two brane configurations with two or three
extra abelian branes which can accommodate the Standard Model with two Higgs
doublets. We analysed the possible hypercharge embeddings and found seven
possible solutions leading to six models (with acceptable string scale $M_S$),
implying the correct charge assignments for all standard model particles.

We further examined the gauge coupling evolution in these models for both,
parallel, as well as intersecting branes and determined the lowest string scale
allowed for all possible alignments of the $U(1)$ branes with respect to the
$U(3)$ and $U(2)$ non-abelian factors of the gauge symmetry. In the parallel
brane scenario, we have identified three models which allow a string scale $M_S$
as low as a few TeV, one model  with string scale of the order $10^6$ GeV and two
models  with high unification scales. Similar results were obtained for the
general case of intersecting branes.

We further analysed the consequences of the third generation fermion mass
relations and in particular $b-\tau$ equality at the string scale on the above
models. In the parallel brane scenario, we found that exact $b-\tau$ Yukawa
unification is obtained only in the model with $M_S\approx 10^3$ TeV, while in
the TeV  string scale models  the $m_b/m_{\tau}$ ratio deviates from unity by
$25\%$. Allowing the $U(1)$ gauge couplings to take arbitrary (perturbative)
values, we found that $b-\tau$ Yukawa unification is possible for a wide string
scale range form $10^6$ up to $10^{12}$ GeV.

{\ack  This research was funded by the program `PYTHAGORAS' (no.\ 1705 project
23) of the Operational Program for Education and Initial Vocational Training of
the Hellenic Ministry of Education under the 3rd Community Support Framework and
the European Social Fund.}

\end{document}